
\magnification 1200

\parskip 3truemm

\def\bi{\bigskip \indent}

\def\br{\hfill\break}

\def\ni{\noindent}
\def\nostrocostrutto#1\over#2{\mathrel{\mathop{\kern 0pt \rlap
  {\raise.2ex\hbox{$#1$}}}
  \lower.9ex\hbox{\kern-.190em $#2$}}}
\def\lsim{\nostrocostrutto < \over \sim}
\def\gsim{\nostrocostrutto > \over \sim}

\def\noi1{noi1}

\def\bi{$\tilde B$}
\def\wi{$\tilde W_3$}
\def\c{\chi}

\def\m{m_{\chi}}

\def\tb{\tan \beta}

\def\apert{A t}

\def\gev{{\rm GeV}}

\null
{
\nopagenumbers
\rightline{DFTT 34/94}
\rightline{August 1994}
\vskip 1truecm

\centerline{\bf SIGNALS OF NEUTRALINO DARK MATTER FROM EARTH AND SUN}

\vskip 1.0truecm
\centerline{A. BOTTINO, N. FORNENGO, G. MIGNOLA}
\baselineskip=13pt
\centerline{\it Dipartimento di Fisica Teorica, Universit\`a di Torino and}
\baselineskip=12pt
\centerline{\it INFN, Sezione di Torino, via P.Giuria 1, 10125 Torino, Italy}
\vskip 0.3truecm
\centerline{and}
\vskip 0.3truecm
\centerline{L. MOSCOSO}
\baselineskip=13pt
\centerline{\it DAPNIA/DSM/SPP}
\baselineskip=12pt
\centerline{\it CE--Saclay, 91191 Gif--sur--Yvette, France}
\baselineskip=12pt

\vskip 3truecm

\centerline{\bf Abstract}

We evaluate the fluxes of  up--going muons detectable in a neutrino
telescope and due to the annihilation of relic neutralinos in the Earth
and in the Sun,
taking realistically into account the fact that neutralinos
might provide only a fraction of the local (solar neighbourhood)
dark matter (DM). We determine the relation between the exposure
$A t$ of the
 neutrino telescope
($A$ being the telescope area, $t$ the live time)
and the
explorable range of the neutralino mass and composition, when
the signal--to--background ratio is optimized by an
appropriate angular selection.

\vfill
\centerline{( to appear in {\it Astroparticle Physics} )}
\bigskip
\bigskip
\bigskip

\footnote{}{
E--mail addresses: (bottino, fornengo, mignola)@to.infn.it \hfill\break
\phantom{E--mail addresses: \ \ \ \ \ \ \ } moscoso@saclay.cea.fr}

\eject
}

\pageno=1
\ni
{\bf 1 \ \ \  Introduction}

\indent
The perspectives of observing, with neutrino telescopes of large area,
indirect signals of relic neutralinos form celestial bodies (Earth and
Sun) are very appealing and are currently the subject of much
investigation [1--3].
The sequence of the physical processes which could produce these
signals are: i) capture of relic neutralinos by the macroscopic
bodies; ii) subsequent  accumulation of these
particles in a region around the centre of these celestial objects;
iii) pair annihilation of the accumulated neutralinos which would
generate, by decay of the particles produced in the  various
annihilation final states, an output of high--energy neutrinos.
These neutrinos may be detected in a neutrino telescope in two ways:
either by looking at events due to the $\nu$--interactions inside the
detector (contained events) or by measuring the muons which are produced
by $\nu_\mu$ and ${\bar \nu}_\mu$ interactions in the rock around the
detector and then traverse it upwardly. In the present paper we only
discuss the latter case; we address the problem of the theoretical
predictions for the fluxes of these up--going muons in a neutrino
telescope and of their discrimination against the
background due to atmospheric neutrinos.
The present analysis aims at improving the previous ones in many
instances:
a) The calculations of the predicted signals for  both
macroscopic bodies (Earth and Sun) are presented in a common
theoretical framework;
in particular, this enables a comparative discussion about the
informations obtainable by measurements of the two
fluxes.
b) It is  realistically taken into account the fact that neutralinos
might provide only a fraction of the local (solar neighbourhood)
dark matter (DM)
density. In order to establish whether and by how much the total DM
density has to be rescaled, the neutralino relic abundance is explicitly
evaluated at any point of the model parameter space where signals are
calculated.
c) It is finally discussed the relation between the exposure
$A t$ of a neutrino telescope
($A$ being the telescope area, $t$ the live time)
and the
explorable range of the neutralino mass (and composition),
taking into account the
optimization of the signal--to--background ratio (S/B).

  The theoretical framework adopted in the present paper is the
one provided by the Minimal Supersymmetric extension of the Standard
Model (MSSM) [4], which, implemented by a general grand unification
relation, may be formulated (for the neutralino sector)
in terms of only three parameters: two
mass parameters $M_2$ and $\mu$, and a third parameter
$\tan \beta$ defined as $\tan \beta = v_u/v_d$ ($v_u$ and
$v_d$ are the v.e.v.'s that give masses to the up--type and down--type
quarks, respectively). The neutralino, $\chi$, is defined to be the
lowest--mass linear combination of the two gauginos (photino and zino)
and the two higgsinos

$$\eqalign{\chi=a_1\tilde\gamma + a_2\tilde Z+a_3\tilde H_1^0
+a_4\tilde H_2^0}. \eqno(1)$$

\noindent
Here $\tilde \gamma$ and $\tilde Z$ are the fields obtained from the
original fields for the U(1) and SU(2) neutral gauginos,
\bi \ and \wi, by a rotation in terms of the Weinberg angle.

It is useful to introduce a composition parameter
$P = {a_1}^2 + {a_2}^2$ which gives the gaugino fractional weight. The
neutralino mass $m_{\chi}$ and the coefficients $a_i$'s depend on the
three model parameters $M_2$, $\mu$ and $\tan \beta$. Since almost
everywhere in the parameter space there is a one--to--one
correspondence: $(M_2, \mu) \leftrightarrow (P, m_{\chi})$, at fixed
$\tan \beta$, in the following we will use as a set of independent
parameters: $P$, $m_{\chi}$ and $\tan \beta$.
This allows us to present
our results in terms of two quantities, $P$ and $m_{\chi}$, which have a
direct physical meaning for the neutralino.
In the following, for $\tan \beta$,
unless differently
specified, we will use the
representative value $\tan \beta = 8$.

We still have to state which values we assign to the unknown
masses of the Higgs bosons and of the sfermions; in fact these masses
enter in the theoretical calculations, since Higgses and sfermions
play a crucial role in the physical processes initiated by
neutralinos.
As for the neutral Higgs bosons, we
recall that in the MSSM there are three neutral Higgs particles:
two CP--even bosons $h$ and $H$ (of masses $m_h$, $m_H$ with $m_H>m_h$)
and a CP--odd one $A$ (of mass $m_A$). Once a value for one of these
masses (say, $m_h$) is assigned, the other two masses ($m_A$, $m_H$) are
derived through mass relationships depending on radiative effects.
In the present paper the Higgs mass $m_h$ is set at the value
of 50 GeV.
As for the sfermion masses, these are set here at the value
$m_{\tilde f} = 1.2 ~ m_\c$, when $m_\c > 45 $ GeV,
$m_{\tilde f} = 45$ GeV otherwise. Only the mass of the stop quark is
assigned a larger value of 1 TeV. The
top mass is fixed at $m_t = 170$~GeV.

\ni
{\bf 2  \ \ \  Neutralino local density}

\indent
An important quantity that enters in the capture rate of relic
neutralinos
by macroscopic bodies is the neutralino local density $\rho_{\c}$.
One can assume that $\rho_{\c}$ is equal to the local value of the
total DM density $\rho_l$, when the neutralino relic
abundance $(\Omega_\chi h^2)$ turns out to be at the level of
an $(\Omega h^2)_{\rm min}$  consistent with $\rho_l$. On the contrary
when $(\Omega_\chi h^2)$ is smaller than $(\Omega h^2)_{\rm min}$ the value
to be assigned to $\rho_{\c}$ has to be appropriately reduced.
Thus we evaluate $\Omega_\chi h^2$  and we determine $\rho_{\c}$ by
adopting a standard rescaling procedure [5]:

$$\eqalign{
\rho_\chi&=\rho_l,
\qquad\qquad~~~ {\rm when}~~~ \Omega_\chi h^2 \geq (\Omega h^2)_{\rm min}
\cr
\rho_\chi&= \rho_l {\Omega_\chi h^2 \over (\Omega h^2)_{\rm min}},
{}~~~{\rm when}~~~ \Omega_\chi h^2 < (\Omega h^2)_{\rm min}
}
\eqno(2)$$

\noindent
Here $(\Omega h^2)_{\rm min}$ is set equal to 0.03.
For $\rho_l$, we have used the value $\rho_l = 0.3$ GeV cm$^{-3}$.

The neutralino relic abundance $ \Omega_\chi h^2$  is evaluated here
following the procedure illustrated in Ref. [6]. By way of example  in
Fig.1 we display $\Omega_\chi h^2$  as a function
of $\m$ for three representative neutralino compositions: i) a
gaugino--dominated composition ($P = 0.9$), ii) a composition of maximal
gaugino--higgsino mixing ($P = 0.5$), iii) a higgsino--dominated composition
($P = 0.1$). As expected, out of the three
compositions displayed in Fig.1, the gaugino--dominated state
provides the largest values of $ \Omega_\chi h^2$.
More substantial values of $ \Omega_\chi h^2$ occurs if one considers
purer
gaugino compositions ($P \gsim 0.99$), as shown in
Fig.2, where, together with the reference value $P = 0.5$, are
also displayed the rather extreme cases: $P = 0.01$ (very pure higgsino
composition) and $P = 0.99$ (very pure gaugino composition). The very
pronounced dips in the plots of Fig.s 1--2 are due to the s--poles in the
$\chi$--$\chi$ annihilation cross section (exchange of the $Z$ and of
the neutral Higgs bosons).

Displayed in Fig.3 are the values of
$ \Omega_\chi h^2$ versus $m_\chi$
in the form of scatter plots obtained by varying $M_2$ and $\mu$ over
a grid of constant spacing in a log--log plane over the ranges
 20 GeV $\leq M_2 \leq$ 6 TeV, 20 GeV $\leq |\mu| \leq$ 3 TeV.
By comparing Fig.3 with Fig.2 one sees that the large spread in
values for $ \Omega_\chi h^2$, displayed in Fig.3,
is due to configurations of extremely pure
composition.

\bigskip

\ni
{\bf 3 \ \ \   Capture rates and annihilation rates}

\indent
The capture rate $C$ of the relic neutralinos by a macroscopic body may
be evaluated by the standard formula [7]

$$C={\rho_{\chi}\over v_{\chi}}\sum_{i}{\sigma_{{\rm el}, i}\over
m_{\chi}m_{i}}(M_{B}f_{i})\langle v^{2}_{esc}\rangle _i
X_i,   \eqno(3)$$

\noindent
where $v_{\chi}$ is the neutralino mean
velocity, $\sigma_{{\rm el}, i}$ is the cross section of the neutralino
elastic scattering off the
nucleus $i$ of mass $m_{i}$,
$M_{B}f_{i}$ is the total mass of the element
$i$ in the body of mass $M_{B}$, $\langle v^{2}_{esc}\rangle_{i}$
is the square escape velocity averaged over the distribution of the
element $i$, $X_{i}$ is a factor which takes account of kinematical
properties occurring in the neutralino--nucleus interactions.
The elastic $\chi$--nucleus cross sections are evaluated according to
the method presented in Ref. [8].

The annihilation rate ${\Gamma}_A$ is expressed in terms of the capture
rate by the formula [9]

$$\eqalign{\Gamma_A={C\over 2} {\rm tanh}^2 ({t\over \tau_A})}\eqno(4)$$

\noindent
where $t$ is the age of the macroscopic body ($t= 4.5~{\rm Gyr}$ for Sun
and Earth),
$\tau_A = (C C_A)^{-1/2}$, and $C_A$ is the annihilation rate per
effective volume of the body, given by

$$\eqalign{C_A={<\sigma v> \over V_0} ({\m \over {20~{\rm GeV}}})^{3/2}}
  \eqno(5)$$

\noindent
$V_0$ is defined as
$V_0=(3 m^2_{Pl} T / (2 \rho \times 10~{\rm GeV}))^{3/2}$
where $T$ and $\rho$ are the central temperature and the central
density of
the celestial body. For the Earth ($T=6000 ~{\rm K}$,
$\rho= 13~{\rm g} \cdot {\rm cm}^{-3}$)
$V_0= 2.3 \times 10^{25} {\rm cm}^3$, for the Sun
($T=1.4 \times 10^7~{\rm K}$,
$\rho= 150~{\rm g} \cdot {\rm cm}^{-3}$)
$V_0= 6.6 \times 10^{28}~{\rm cm}^3$.
$\sigma$ is the neutralino--neutralino annihilation cross
section and $v$ is the relative velocity.
$<\sigma v>$ is calculated with all the contributions
at the tree level as in Ref.[6],
with the further inclusion here of the two gluon
annihilation final state [10].

We recall that, according to Eq.(4), in a given macroscopic body the
equilibrium between capture and annihilation ({\it i.e.}
$\Gamma_A \sim C/2$ ) is established
only when $t \gsim \tau_A$.
It is worth noticing that the neutralino density $\rho_{\c}$, evaluated
according to Eq.(2), enters not only in $C$ but also in $\tau_A$
(through $C$).
Therefore the use of a correct value for $\rho_{\c}$ (rescaled according
to Eq.(2), when necessary) is important also in determining
whether or not the equilibrium is already set in a macroscopic body.

 In Fig.4 we give the results of our calculations for
${\rm tanh}^2(t/\tau_A)$ and for $\Gamma_A$ in the case of
the Earth for three representative $\c$--compositions:
$P = 0.1, 0.5, 0.9$.
For simplicity, in these figures, as well as in the
 following ones, only the results concerning positive values of the
 parameter $\mu$ are shown; similar results hold for negative values of
$\mu$. The upper part of Fig.4 shows that
the  equilibrium between capture and annihilation is  not reached for
$m_\c \gsim m_W$; thus a substantial suppression occurs in $\Gamma_A$
at large values of $m_\chi$ because of the factor
${\rm tanh}^2(t/\tau_A)$. In Fig.5  the two quantities
${\rm tanh}^2(t/\tau_A)$ and $\Gamma_A$ are plotted for other
representative values of neutralino compositions: $P = 0.01, 0.99$
(the value $P = 0.5$ is shown again for comparison).
In Fig.s 4--5 we see that, for $m_\chi \lsim 70~$GeV, $\Gamma_A$
shows the characteristic bumps due to fact that the capture is very
efficient when the neutralino mass matches the nuclear mass of some
important chemical component of the Earth (O, Si, Mg, Fe). Superimposed
to these properties are the features due to the rescaling of $\rho_\chi$.
In fact this rescaling implies here both a general suppression and the
appearance of the peculiar dips reminiscent of the singularities in the
annihilation cross section (discussed in Sect. 2). For comparison with
Fig.4 we show in Fig.6 how the two quantities
${\rm tanh}^2(t/\tau_A)$ and $\Gamma_A$  would appear, if the rescaling
were not applied.

Whereas for the Earth the equilibrium condition depends sensitively on the
values of the model parameters, in the case of the Sun equilibrium
between capture and annihilation is reached for the whole range of
$m_\chi$, due to the much more efficient capture rate implied by the
stronger gravitational field [7]. $\Gamma_A$ for the Sun is
shown in Fig.7.

What we have defined above is the annihilation  rate referred to a
macroscopic body as a whole. This is certainly enough for the Sun which
appears to us as a point source. On the contrary, in the case of the
Earth, one has also to define an annihilation rate referred to a unit
volume at point $ \vec r$ from the Earth center

$$ \Gamma_A (r)= {1 \over 2} <\sigma v> n^2(r) \eqno(6)$$

\noindent
where $n(r)$ is the neutralino spatial density which may be written
as [9]

$$n(r) = n_0 {\rm e}^{-\alpha m_\c r^2};\eqno(7)$$

\noindent
here $\alpha = 2 \pi G \rho / (3 T)$ and $n_0$ is such that

$$\Gamma_A = {1 \over 2} <\sigma v> \int d^3r\, n^2(r)\, . \eqno(8)$$

\bigskip

\ni
{\bf 4 \ \ \    Neutrino fluxes}

\indent
The differential neutrino flux due to $\c$--$\c$ annihilation
in a distant source (Sun) is given by

$$\eqalign{{dN_{\nu}\over dE_{\nu}}={{{\Gamma}_A}\over 4\pi d^{2}}\sum_{F,f}
B^{(F)}_{\c f}{dN_{f \nu}\over dE_{\nu}}} \eqno(9)$$

\noindent
where $d$ is the distance from the source, $F$ denotes the
$\c$--$\c$ annihilation final states,
$B^{(F)}_{\c f}$ denotes the branching ratios into
heavy quarks, $\tau$ lepton and gluons in the channel $F$;
$dN_{f \nu}/dE_{\nu}$ is the differential distribution
of the neutrinos generated by the hadronization of quarks
and gluons and the subsequent hadronic semileptonic decays.

    In the case of the Earth we define a differential neutrino flux
generated at a distance $R$ by the $\c$--$\c$ annihilations
which take place at a point $\vec r$
from the center of the Earth

$$\eqalign{{dN_{\nu}\over dE_{\nu} d^3 r}
={{<\sigma v> n^2(r)}\over
8\pi R^{2}}\sum_{F,f}
B^{(F)}_{\c f}{dN_{f \nu}\over dE_{\nu}}} \eqno(10)$$

\noindent
This is the neutrino spectrum which is used in our Monte Carlo
code to evaluate
the flux of up--going muons, as illustrated in Sect. 5.

However, it is also useful to write the neutrino differential flux
at a point on the Earth surface, once the integration over the
distribution $n(r)$ is performed, at fixed nadir angle $\theta$
[7]

$$\eqalign{{d^2 N_{\nu}\over {dE_{\nu} d\cos \theta}}=G(\theta)
{{\Gamma_A}\over
4\pi R^{2}_\oplus}\sum_{F,f}
B^{(F)}_{\c f}{dN_{f \nu}\over dE_{\nu}}}. \eqno(11)$$

\ni $G(\theta)$ is given by

$$G(\theta) \simeq 2 (2 m \beta) {\rm e}^{-2 m \beta \sin^2 \theta}
\eqno(12)$$

\ni where $\beta =\alpha R^2_\oplus$, for $m_\c \gsim 10$ GeV
and $\theta \lsim
60^\circ$. This flux, as all other fluxes discussed afterwards,
refer to surfaces orthogonal to the incoming direction.

As far as the differential neutrino spectra are concerned,
we briefly quote the
relevant ingredients that we used in the calculation.
A Jetset 7.2 Monte Carlo [11]
code was used to compute the
neutrino spectra due to b and c quarks, tau lepton and gluons.
Following Ref. [12]
we have neglected the contributions of the light quarks directly
produced in the annihilation process or in
the hadronization of heavy quarks and gluons,
because these light particles stop inside the medium
(Sun or Earth) before their
decay.
For the case of the Sun
also the energy loss of the heavy hadrons
in the solar medium was considered.
The spectra due to heavier final states,
{\it i.e.} Higgs bosons, gauge bosons and t
quark, were computed analytically by following the
decay chain down to the production
of a b quark, c quark or a tau lepton;
the result of the Monte Carlo
was used to obtain the final neutrino output.
Because of the high column density of the solar medium,
the absorbtion and the energy loss of the
produced neutrinos were also included.

As an example, we show in Fig.8 the neutrino spectra for neutralino
annihilation in the Earth, for $P = 0.5$ and for three different
values of the neutralino mass: $\m =$ 20, 40 and 120 GeV.
The different shapes depend on which final states dominate in
the annihilation cross section. For instance the spectrum at
$\m =$ 120 GeV shows the typical features due to the opening of the
$W^+$--$W^-$ final state in the $\chi$--$\chi$ annihilation.
The relative magnitudes of the spectra at different values of
$m_\chi$ are driven by the dependence of the annihilation rate
$\Gamma_A$ on the model parameters.

\bigskip

\ni
{\bf 5 \ \ \    Signals of up--going muons}

\indent
Let us turn now to the calculation of the various distributions and
rates for the up--going muons generated by the neutrino fluxes of
Eq.s(9--10) (signals) and by the neutrinos produced in the Earth
atmosphere (background).
The conversion of the muon neutrinos into muons and the subsequent muon
propagation up to the detector have been treated
using a Monte Carlo simulation.

The muon flux at the detector depends on the following quantities:
the neutrino energy spectrum and spatial distribution,
the differential $\nu_{\mu} - N$ charged current cross section for
the muon production and
the energy losses and the multiple scattering of the muon in the
rock surrounding the detector.

As for the signals, the neutrino energies
have been generated
in the range 2 GeV  $\leq E_{\nu} \leq \m$
according to the distributions of Eq.(9) for the Sun and of
Eq.(10) for the Earth. We emphasize that in the latter case
the neutrino production point has been generated in a region
around the Earth center with a gaussian distribution
$n^2 (r)$, where $n(r)$ is given in Eq.(7).
The atmospheric neutrinos have been generated upwardly
according to the differential distribution given by [13]
for neutrinos with $E_{\nu} \leq $ 100~GeV
and by [14]
for neutrinos with $E_{\nu} > $ 100~GeV.

In the cross section for the charged current $\nu_{\mu} - N$
deep inelastic scattering
we have employed the
CTEQ parton distribution functions [15].
The energy losses and multiple scattering of the muon
have been calculated by propagating the muon in standard
rock by steps of $10^4$~ g~ cm$^{-2}$ when the muon energy was larger than
30 GeV, of 500~ g~ cm$^{-2}$ when it was between 1.2 and 30~ GeV, and equal
to $E_{\mu}/(2.4\cdot 10^{-3}~{\rm GeV})$~ g~ cm$^{-2}$ for the
final step when $E_{\mu} < 1.2$~ GeV.
The residual energy and the muon direction were kept in memory for
each step from the muon production to the muon stopping.
The detector entry point has been randomly generated with a uniform
probability along the whole muon track. The muon energy and direction
at the entry point has been calculated by interpolating between the
initial and final values for the step where the entry point occurred.
A random angle has been included to take into account a
detector angular resolution of $2^\circ $.

\bigskip
\vfill
\eject

\ni
{\bf 6 \ \ \    Results  and conclusions}

\indent
We show in Fig.s 9--11 a sample of our results for the signals expected
from the Earth at the usual representative point $\tan \beta = 8$ (for the
fluxes and the angular distributions shown in these figures and in the
following ones the threshold for the muon energy is set at the value
$E^{\rm th}_{\mu} = 2~\gev$).
Fig.9 displays the integral muon flux versus $m_{\chi}$
for various neutralino compositions. This has been obtained with a
no--straggling approximation [16], $\it i.e.$ the muons are produced and
propagated in the forward direction (when this approximation is applied,
 then Eq.(11) for the neutrino distribution is useful).
The general structure of the flux clearly reflects the
features of
the annihilation rate in Earth, previously discussed. The horizontal
line represents the Kamiokande upper bound
for signals coming from the Earth [2]:
$4.0 \cdot 10^{-14} {\rm cm}^{-2} {\rm s}^{-1}$ (90 \% C.L.);
this value refers to a half-aperture of $\theta_t = 30^\circ$ for a
cone centered at the nadir.
By comparing this upper limit with our fluxes we see that
the regions explored by Kamiokande (at our representative point:
tan $\beta = 8, m_h$ and $m_{\tilde f}$ chosen as stated in Sect. 1)
concern the neutralino mass range
$50~\gev \lsim m_\c \lsim 80~\gev$, for $0.1 \lsim P \lsim 0.9$;
the mass range is narrower for purer neutralino compositions.

By way of example we show in Fig.s 10--11
the differential spectra and the angular distributions
for the case $P = 0.5$ and $m_\chi = 120$ GeV.
In Fig.10 the spectra obtained with the MC code  for the expected signal
and for the background are displayed by the two histograms. The solid
line represents the spectrum for the signal obtained with the
no--straggling approximation; it is clear from the plot that
this approximation is quite good for the
spectrum, and then also for the integrated flux. The spectra shown in
this figure refer to the muons entering the detector within a cone of
half--aperture $\theta_t = 30^\circ$ centered at the nadir.
In Fig.11 the angular distribution due to the signal, markedly peaked
around the nadir direction, is compared with the almost flat background
distribution.

Now a very relevant question is what is the minimal exposure $A t$
necessary to provide a good S/B discrimination. To put it into a more
quantitative way, we define  an $(A t)_{\rm min}$ as the minimal exposure
necessary to see a $4 \sigma$ effect (with a signal of at least 4
events) [17]. $(A t)_{\rm min}$ is displayed in
Fig.12a (the angle $\theta_t$ of half-aperture from the nadir is again
set at the value $\theta_t = 30^\circ$ here). However, it is important
to remark that, due to the different features of the angular
distributions for the signal and for the background (illustrated for
instance in Fig.11), $(A t)_{\rm min}$ may be optimized by appropriately
choosing the angular window, $i.e. ~\theta_t$. We call
$(A t)_{\rm opt}$ the least value of $(A t)_{\rm min}
$ when $\theta_t$ is varied,
and we denote by $\theta_{\rm opt}$
the corresponding value of $\theta_t$.
This optimization procedure has to be applied to $(A t)_{\rm min}$
for each
value of the integral flux $\Phi_{\mu}$. For the case illustrated in
Fig.11 we find $\theta_{\rm opt} \simeq 5^\circ$,
$(A t)_{\rm opt} \simeq 3020$ m$^2$ yr.
Fig.12b displays the dependence of $(A t)_{\rm opt}$ on $m_{\chi}$ and
$P$ for our usual representative case.
To obtain the result of Fig.12b the optimization, based on the angular
distribution, has been performed
by using the results of the MC simulations at 4 values of the neutralino
mass: $m_\chi = $ 20, 40, 60, 120 GeV,
by interpolating between these values
and by conservatively using for $\m > 120$ GeV the same angular window
as for $\m = 120$ GeV.
For the integral flux we
employed the results (shown in Fig.9a) obtained with the no--straggling
approximation.
By comparing Fig.s 12a and 12b one sees that the
optimization improves significantly the experimental capabilities
especially at large $m_\chi$ values; in fact for these values the
extension of the annihilation region is very much concentrated around
the Earth center.
Fig.12b shows realistically by which amount the explorable ranges of the two
neutralino parameters $m_\chi$ and $P$ increase as the exposure
$A t$ is increased. We remind that the largest value for $A t$
obtained up to now is the one of Kamioka: $A t = 770$ m$^2$ yr [2].
A sizeable improvement in $A t$ will be provided by MACRO [18].

Let us turn now to the signals from the Sun. In Fig.13 the relevant flux of
up--going muons is plotted versus $m_\chi$ in the case of the
representative point under discussion; $\Phi_\mu$ is the flux through
a surface orthogonal to the Sun direction.
The horizontal line denotes the present upper bound of
$6.6 \cdot 10^{-14} {\rm cm}^{-2} {\rm s}^{-1}$ (90 \% C.L.)
(an half-aperture $\theta = 25^\circ$ was taken here for a cone centered
at the Sun direction) [2].
It is clear that this limit does not introduce any constraint here.
In Fig.14 the corresponding
$(\epsilon A t)$--$m_\chi$ plot is displayed (a
cone of $5^\circ$ half--aperture
around the Sun direction has been used for the evaluation of the
background, $\epsilon$ is the on-source duty factor). In this
plot we see that, in order to start some
exploration about relic neutralinos using the flux from the Sun, one
needs at least $(\epsilon A t) \simeq 10^4$ m$^2$ yr
(this could provide information in the range 70 GeV$ \lsim m_\chi
\lsim $ 150 GeV). However, it is remarkable that by slightly increasing
$(\epsilon A t$) above
this level, the explorable range in terms of $P$ and $m_\chi$ expands
quite rapidly towards the high values of the neutralino mass.

An interesting point concerns the relative importance of the up-going
muon fluxes from the Earth and from the Sun. As we see from
Fig.s (9,13), for light neutralinos the signal from the Sun is much
smaller than the signal from the Earth; as $m_\chi$ increases the signal
from the Sun may overcome the other, since the Sun gravitational field
is much more efficient in capturing the neutralinos.
More relevant for a comparative discussion about the signals from the
two sources are the two diagrams of the exposure versus $m_\chi$
displayed in Fig.s 12b, 14. A very appealing
situation is the one where the signals from both sources could be
detected. For instance, assuming $\epsilon \simeq 0.5$, an exposure
$A t \simeq 10^5$ m$^2$ yr would be sensitive to both signals in the
range 35 GeV $\lsim m_{\chi} \lsim$
 200 GeV.

A word of warning is in order here to remind that the present
evaluations depend sensitively on a number of free parameters. As far as
$\tan \beta$ is concerned, we may notice that the dependence of the
signals on this parameter is rather involved, but, as a general trend,
the expected signals increase with the value of $\tan \beta$.
For instance, for the Sun, moving from the value
$\tb = 8$ to the value $\tb = 20$ the flux is enhanced by about a factor
5 for $\m \gsim 200$ GeV; but, if the value of $\tan \beta$ is reduced
from $\tb = 8$ to $\tb = 2$, the flux is suppressed
(for $m_\chi \gsim$ 50 GeV) roughly by an order of magnitude.
The quantities calculated in the present paper are also rather sensitive
on the values employed for the Higgs and the sfermion masses. For
instance, our signals for the Earth are roughly proportional to
$m_h^{-4}$. In the present paper we have used for the
Higgs and the sfermion masses values close to their present lower
bounds. This choice enables one to establish, by use of the figures
previously reported, what is the minimum exposure necessary for a
detector to be sensitive to the Earth and/or Sun signals and how the
exploration into the DM neutralino properties may be progressively
improved by expanding the size of the experimental apparatus.

To summarize we may conclude that the exploration in the realm of the relic
neutralinos by neutrino telescopes, after the first insights provided by
Kamiokande [2] about the Earth signals, will certainly be significantly
developed by the underwater (under--ice) detectors under
construction [3]. In order to have good perspectives for a systematic
investigation of this field one should finally aim at an apparatus of
very large area of order of $10^6$ m$^2$.

\bigskip

\ni
{\bf  \ \ \  Acknowledgments}

\indent
We are grateful to professor V. S. Berezinsky for very interesting discussions.

\vfill
\eject

\ni
{\bf References}
\indent

\item{[1]}
G.F. Giudice and E. Roulet, {\it Nucl. Phys.} {\bf B316}
(1989) 429; \hfill \break
G.B. Gelmini, P. Gondolo and E. Roulet, {\it Nucl. Phys.}
{\bf B351} (1991) 623; \hfill \break
M. Kamionkowski,  {\it Phys. Rev.} {\bf D44} (1991) 3021; \hfill \break
A. Bottino, V. de Alfaro, N. Fornengo, G. Mignola and
M. Pignone, {\it Phys. Lett.} {\bf B265} (1991) 57; \hfill \break
F. Halzen, M. Kamionkowski and T. Steltzer, {\it Phys. Rev.} {\bf D45}
(1992) 4439; \hfill \break
V.S. Berezinsky, {\it Nucl. Phys.} (Proc. Suppl.) {\bf B31} (1993) 413
(Proc. Neutrino 92, Ed. A. Morales) \hfill\break
M. Mori et al., {\it Phys. Rev.} {\bf D48} (1993) 5505; \br
M. Drees, G. Jungman, M. Kamionkowski and M.M. Nojiri,
{\it Phys. Rev.} {\bf D49}
(1994) 636; \hfill\break
R. Gandhi, J.L. Lopez, D.V. Nanopoulos, K. Yuan and A. Zichichi,
{\it Phys. Rev.} {\bf D49} (1994) 3691.

\item{[2]}
M. Mori et al. (Kamiokande Collaboration),
{\it Phys. Lett.} {\bf B289} (1992) 463.

\item{[3]}
J.G. Learned, {\it Nucl. Phys.} (Proc. Suppl.)
{\bf B31} (1993) 456 (Proc. Neutrino 92, Ed. A. Morales); \hfill \break
F. Halzen, {\it Proc.
Fourth International Workshop on Neutrino Telescopes}
(1992), Ed. M. Baldo Ceolin; \hfill \break
G.V. Domogatsky,
{\it Nucl. Phys.} (Proc. Suppl.) {\bf B35} (1994) 290 (Proc. TAUP 93,
Ed.s C.Arpesella, E.Bellotti, A.Bottino); \hfill\break
L.K. Resvanis, {\it loc. cit.} 294. \hfill\break

\item{[4]}
H.P. Nilles, {\it Phys. Rep.} {\bf 110} (1984) 1; \hfill \break
H.E. Haber and G.L. Kane, {\it Phys. Rep.} {\bf 117} (1985) 75; \br
R. Barbieri, {\it Rivista Nuovo Cimento} {\bf 11} (1988) 1. \br

\item{[5]}
T.K. Gaisser, G. Steigman and S. Tilav, {\it Phys. Rev.}
 {\bf D34}(1986)2206.

\item{[6]}
A. Bottino, V.de Alfaro, N. Fornengo, G. Mignola and M. Pignone,
{\it Astroparticle Physics}
{\bf 2} (1994) 67.

\item{[7]}
A. Gould, {\it Ap. J.} {\bf 321} (1987) 571;
{\it Ap. J.} {\bf 328} (1988) 919;
{\it Ap. J.} {\bf 368} (1991) 610.

\item{[8]}
A. Bottino, V. de Alfaro, N. Fornengo, A. Morales, J. Puimed\'on and
S. Scopel, {\it Mod. Phys. Lett.} {\bf A7} (1992) 733.
Notice that in the present paper
for the Higgs boson--nucleon coupling we use
the evaluation of
J. Gasser, H. Leutwyler and M. E. Sainio,
{\it Phys. Lett.} {\bf B253} (1991) 252.

\item{[9]}
K. Griest and D. Seckel,  {\it Nucl. Phys.} {\bf B283} (1987) 681.

\item{[10]}
M. Drees, G. Jungman, M. Kamionkowski and M.M. Nojiri,
{\it Phys. Rev.} {\bf D49} (1994) 636.

\item{[11]}
T. Sj\"ostrand, {\it Comp. Phys. Comm.} {\bf 39} (1986) 347;
{\it Comp. Phys. Comm.} {\bf 43} (1987) 367;
CERN--TH 6488/92.

\item{[12]}
S.Ritz and D.Seckel, {\it Nucl. Phys.} {\bf B304} (1988) 877.

\item{[13]}
G. Barr, T.K. Gaisser and T. Stanev, {\it Phys. Rev.} {\bf D39} (1989) 3532
and private communication.

\item{[14]}
L.V. Volkova, {\it Sov. J. Nucl. Phys.} {\bf 31} (1980) 784.

\item{[15]}
J. Botts, J.G Morfin, J.F. Owens, J. Qiu, W. Tung and H. Weerts,
{\it Phys. Lett.} {\bf B304} (1993) 159.

\item{[16]}
T.K.Gaisser: {\it Cosmic Rays and Particle Physics}
(Cambridge University Press, 1990).

\item{[17]}
A more refined statistical treatment of the S/B
discrimination will be presented by the present authors in a forthcoming
paper.

\item{[18]}
D.G.Michael (MACRO Collaboration),
{\it Nucl. Phys.} (Proc. Suppl.) {\bf B35} (1994) 235 (Proc. TAUP 93,
Ed.s C.Arpesella, E.Bellotti, A.Bottino). \hfill\break

\vfill
\eject

\ni
{\bf Figure Captions}

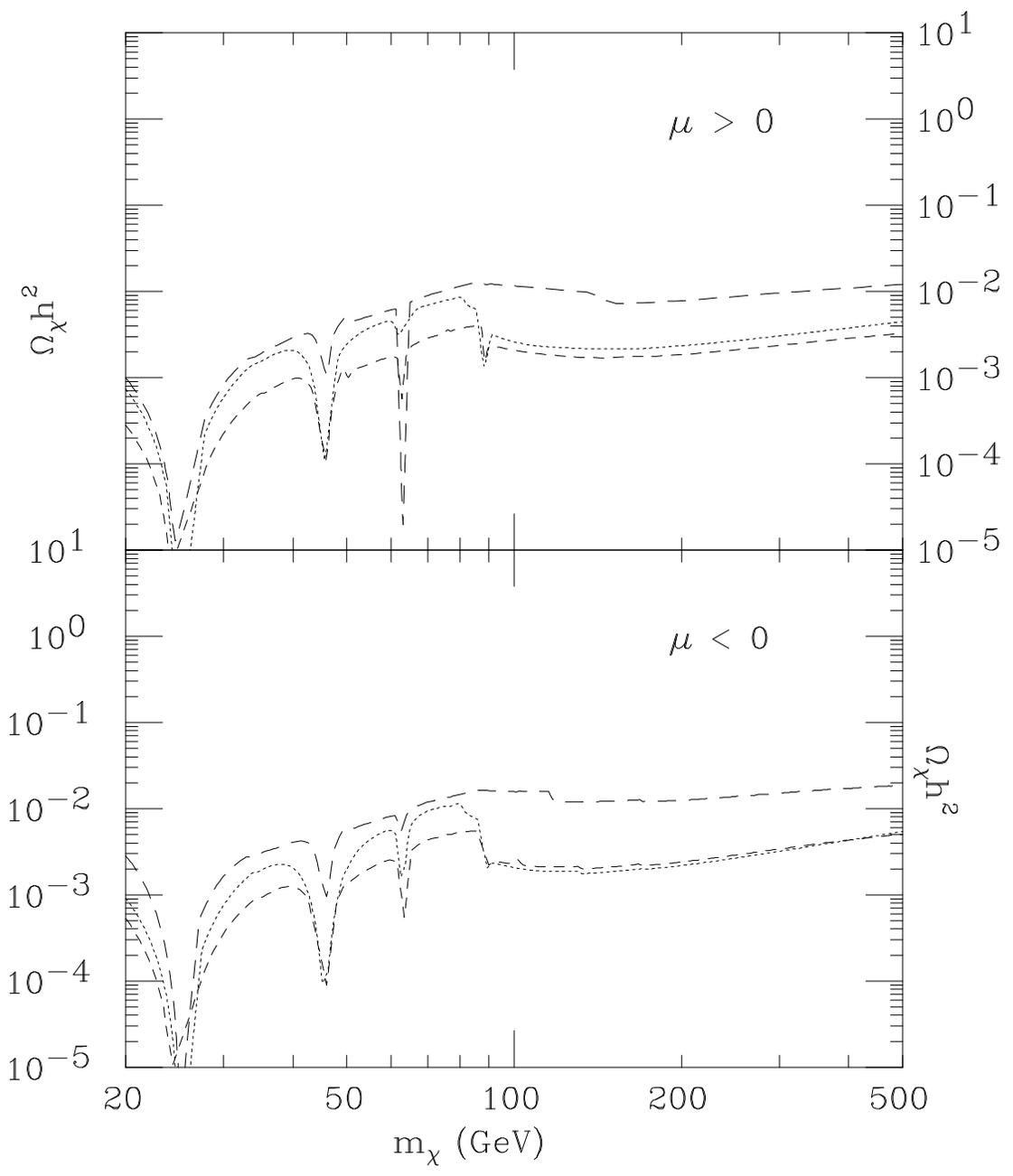
{\bf Figure 1} Neutralino
relic abundance  $ \Omega_\chi h^2$  as a function of $m_\c$ for
three different neutralino compositions $P = 0.1$ (dotted line),
$0.5$ (short--dashed line), $0.9$ (long--dashed line). The upper plot
refers to positive values of $\mu$, the lower to negative values.

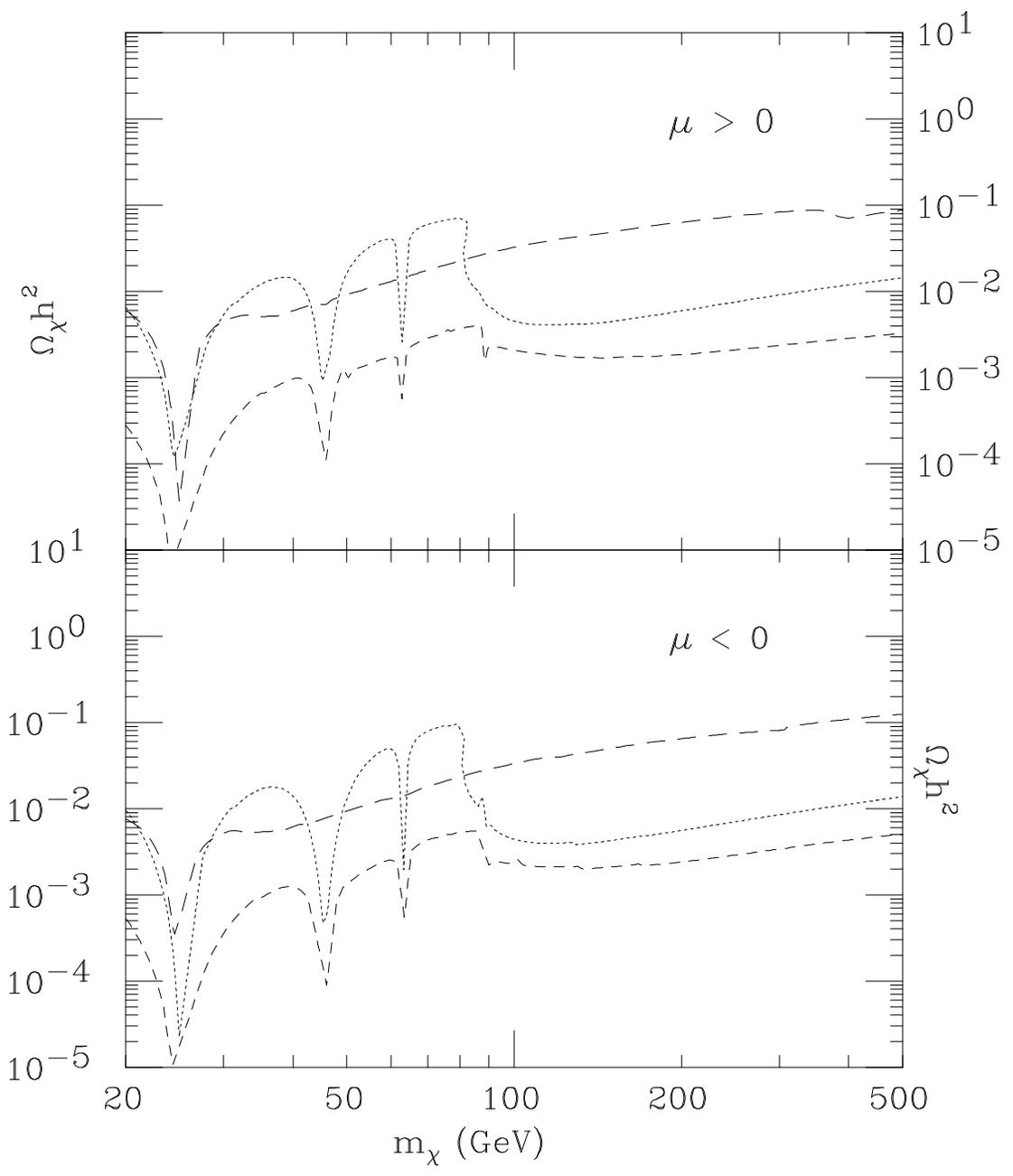
{\bf Figure 2} Same as in Fig.1, except that now composition are:
$P = 0.01$ (dotted line),
$0.5$ (short--dashed line), $0.99$ (long--dashed line).

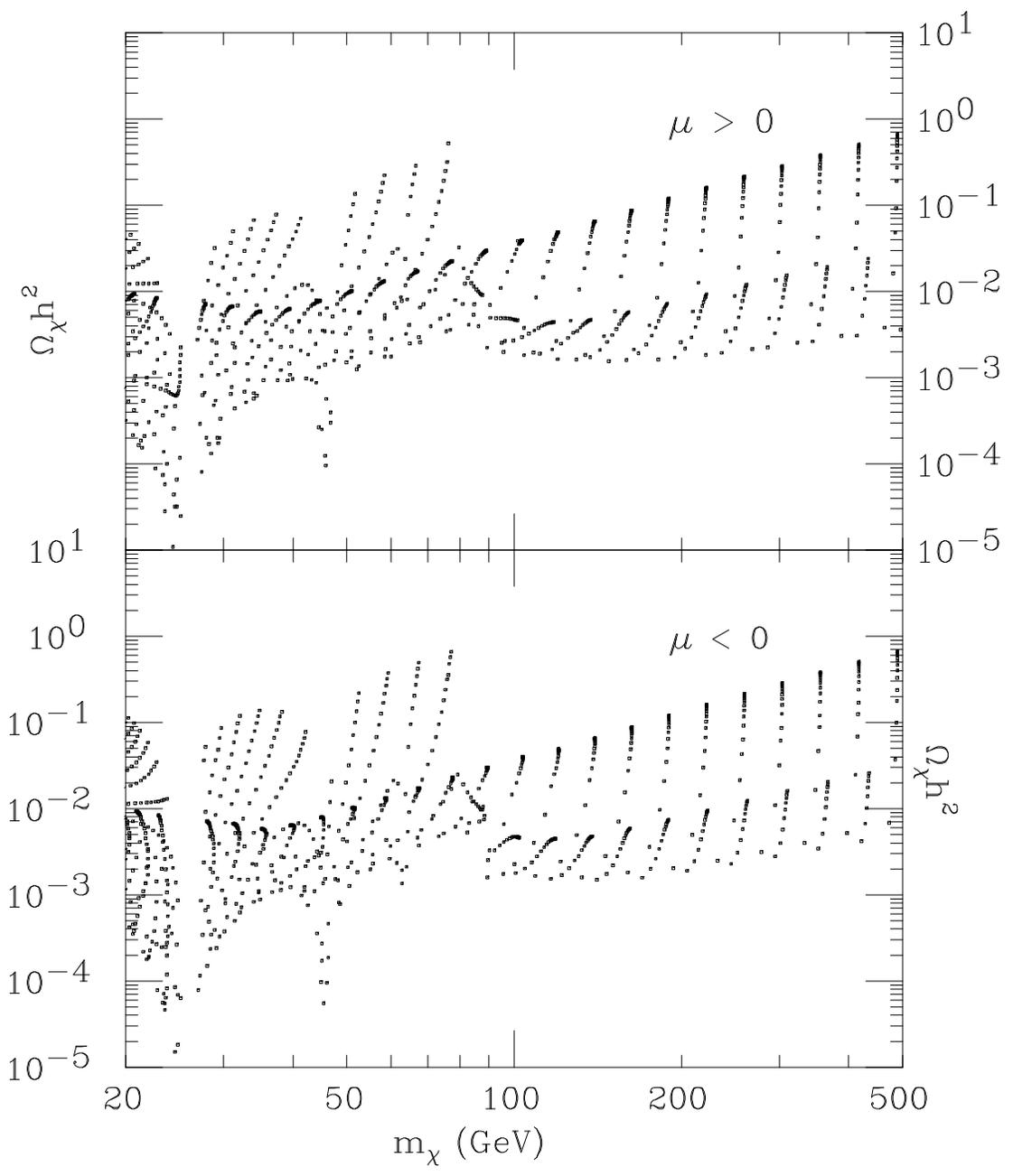
{\bf Figure 3} Scatter plot for  the neutralino
relic abundance $\Omega_\chi h^2$  as a function of $m_\c$.
The two MSSM mass parameters $M_2, \mu$ are varied in the ranges:
20 GeV $\leq |\mu| \leq$ 3 TeV, 20 GeV $\leq M_2 \leq$ 6 TeV.

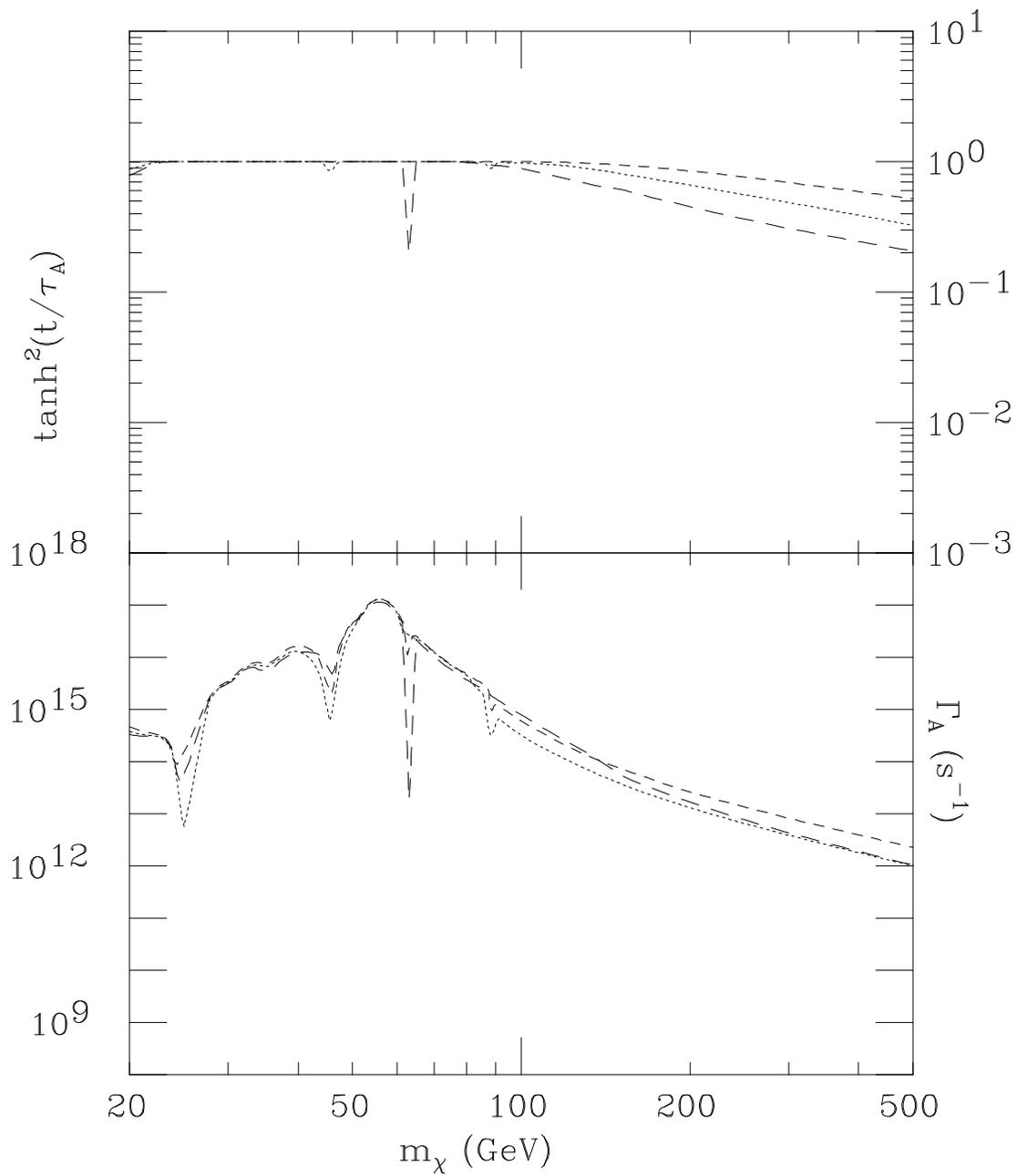
{\bf Figure 4}
${\rm tanh}^2(t/\tau_A)$ and $\Gamma_A$ for the Earth
are given as functions of $\m$, for the
three representative neutralino compositions $P = 0.1$ (dotted line),
$0.5$ (short--dashed line), $0.9$ (long--dashed line).

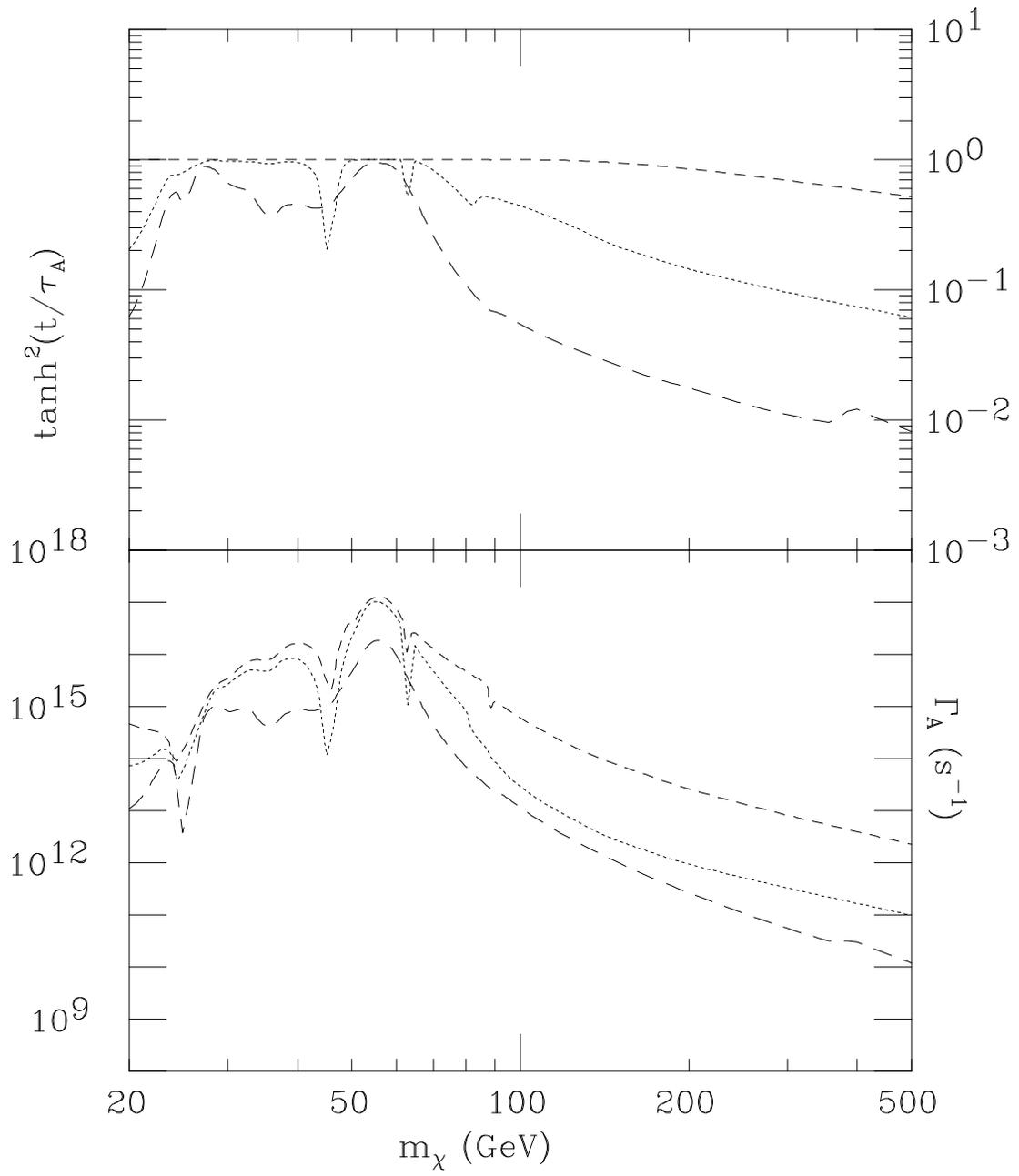
{\bf Figure 5}
Same as in Fig.4, except that now composition are:
$P = 0.01$ (dotted line),
$0.5$ (short--dashed line), $0.99$ (long--dashed line).

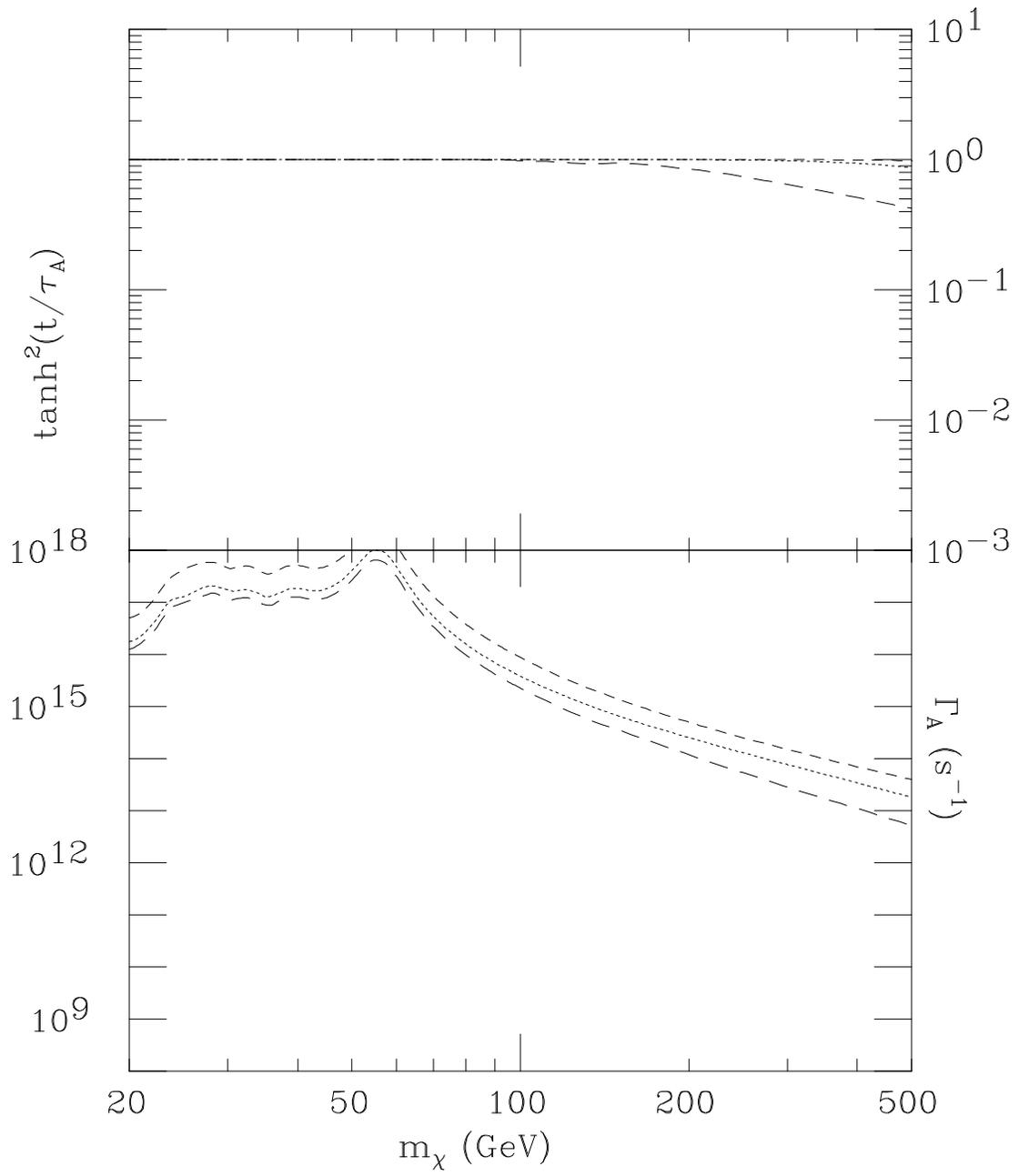
{\bf Figure 6}
Same as in Fig.4, except that now the neutralino  local density
$\rho_{\chi}$ is not rescaled.

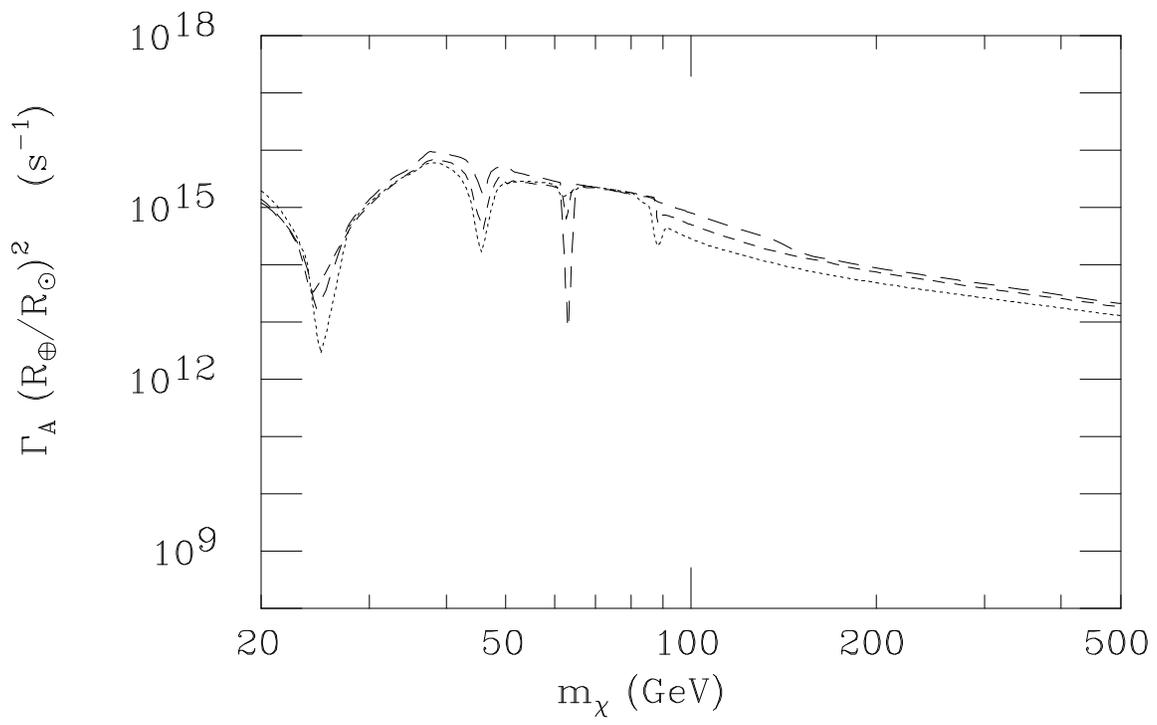
{\bf Figure 7}
Annihilation rate $\Gamma_A$ for the Sun
as a functions of $\m$, for the
three representative neutralino compositions $P = 0.1$ (dotted line),
$0.5$ (short--dashed line), $0.9$ (long--dashed line).
To easily compare the annihilation rate for the Earth to the one
for the Sun, the plot shown here for the Sun refers to the effective
annihilation rate, defined as $\Gamma_A$ times the ratio
$(R_\oplus/R_\odot)^2 = 1.8 \times 10^{-9}$, where $R_\oplus$ is
the radius of the Earth and $R_\odot$ is the Sun-Earth distance.

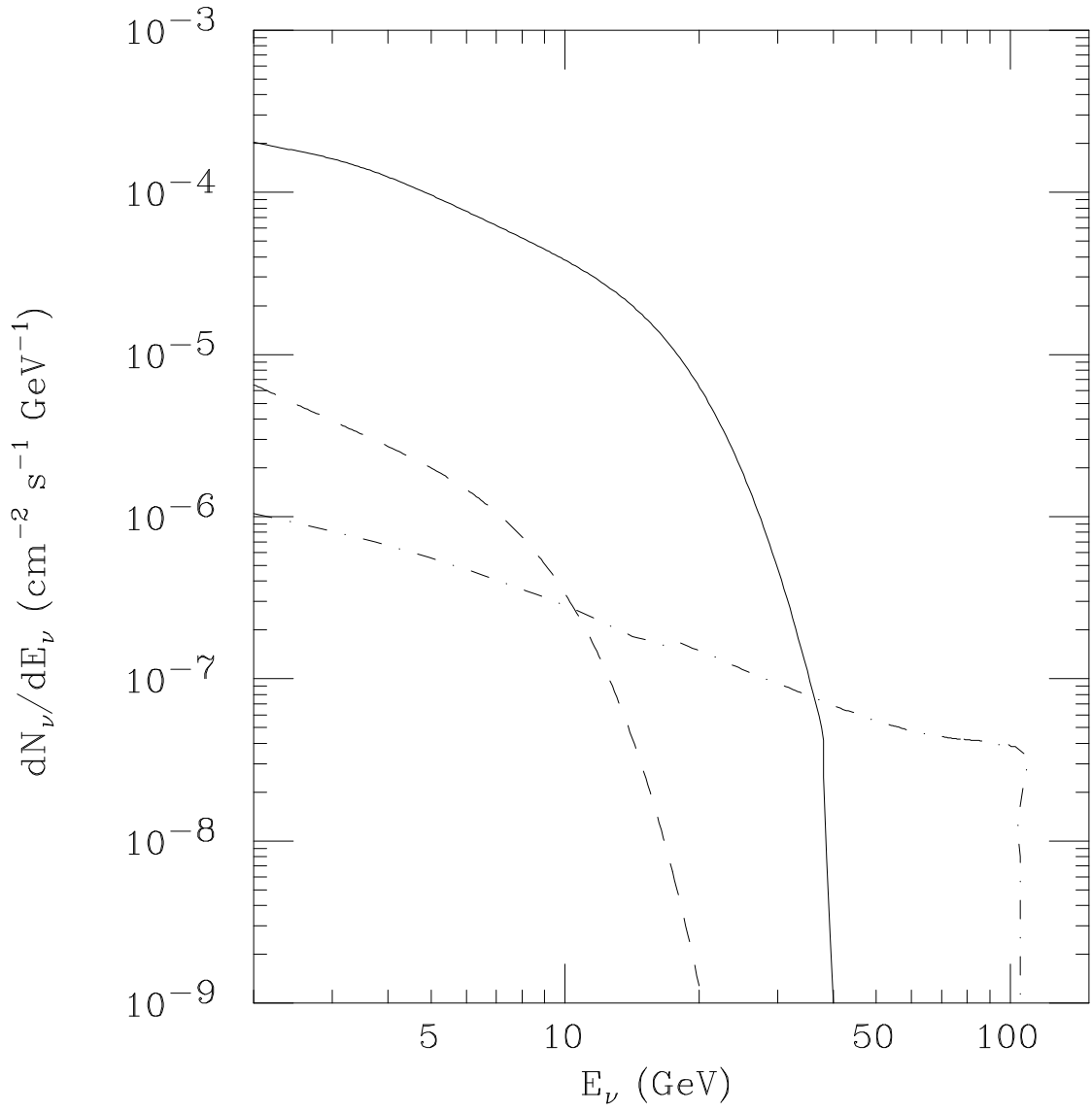
{\bf Figure 8}
Differential neutrino spectra $dN_\nu / dE_\nu$ for
neutralino annihilation in the Earth, as a function
of the neutrino energy $E_\nu$. These spectra
are calculated for a neutralino composition
$P = 0.5$ and for three different values of
neutralino mass: $\m =$ 20 GeV (dashed line),
$\m =$ 40 GeV (solid line) and $\m =$ 120 GeV (dash--dotted line).

{\bf Figure 9}
Flux $\Phi_\mu$ of the up--going muons as functions of $m_\c$
for $\c$--$\c$ annihilation in the Earth.
The threshold for the muon energy is $E^{\rm th}_{\mu} = 2~\gev$.
In figure (a) the three representative neutralino compositions are
$P = 0.1$ (dotted line), $0.5$ (short--dashed line),
$0.9$ (long--dashed line).
In figure (b) $P = 0.01$ (dotted line),
$0.5$ (short--dashed line), $0.99$ (long--dashed line).
The horizontal line in figure (a)
represents the Kamiokande upper bound
$4.0 \cdot 10^{-14} {\rm cm}^{-2} {\rm s}^{-1}$ (90 \% C.L.) [2].

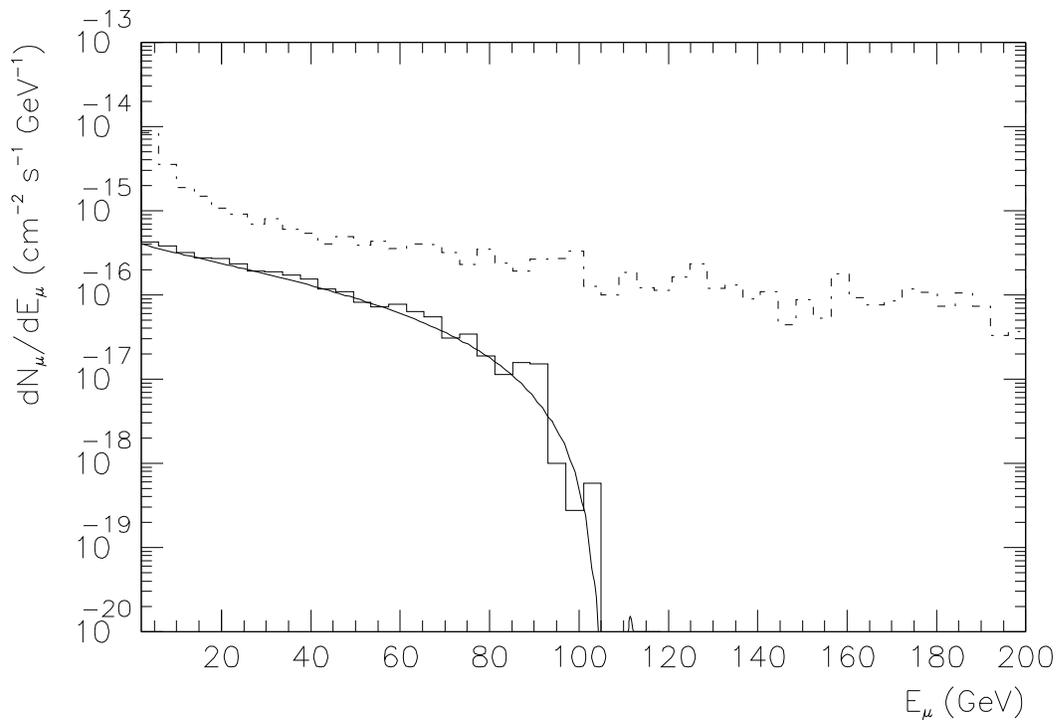
{\bf Figure 10}
Differential muon spectra $dN_\mu / dE_\mu$ as a function
of the muon energy $E_\mu$.
The solid histogram is the muon spectrum obtained with our
MC simulation for
$P=0.5$ and $\m=$ 120 GeV. The solid line
represents the muon spectrum calculated in the no--straggling
approximation. The dash--dotted histogram is the muon
energy distribution for the background of atmospheric neutrinos.
The spectra refer to muons entering the detector within a cone
of half--aperture of $30^\circ$ centered at the nadir.

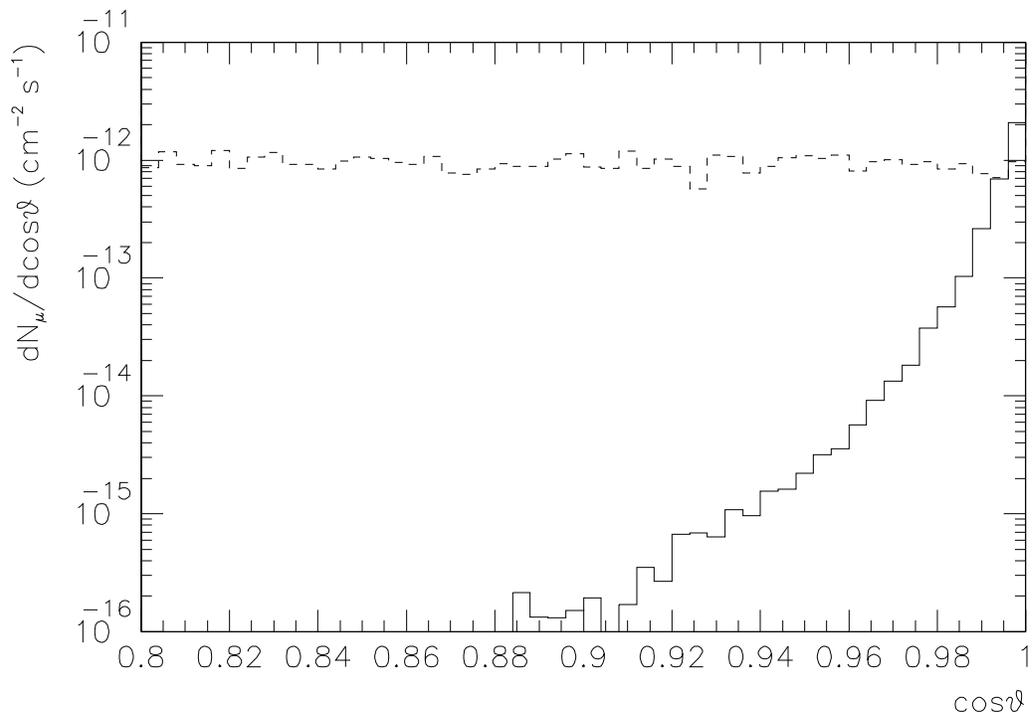
{\bf Figure 11}
Muon angular distribution $dN_\mu / d\cos\theta$ as a function
of $\cos\theta$, where $\theta$ is the nadir angle.
The solid histogram is the angular distribution of the signal
obtained with our
MC simulation for $P=0.5$ and $\m=$ 120 GeV.
The dashed histogram is the muon
angular distribution for the background of atmospheric neutrinos.

{\bf Figure 12}
Exposure necessary to have a $4 \sigma$ effect
with a signal of at least 4 events,
as a function of neutralino mass $\m$.
The figure refers to
a signal coming from the Earth.
The three representative neutralino compositions are
$P = 0.1$ (dotted line),
$0.5$ (short--dashed line), $0.9$ (long--dashed line).
In figure (a) $\apert_{\rm min}$ is displayed keeping
the half--aperture angle of the detector fixed at $30^\circ$ from
the center of the Earth.
Figure (b) displays the exposure $\apert_{\rm opt}$ obtained by
employing the optimization procedure.

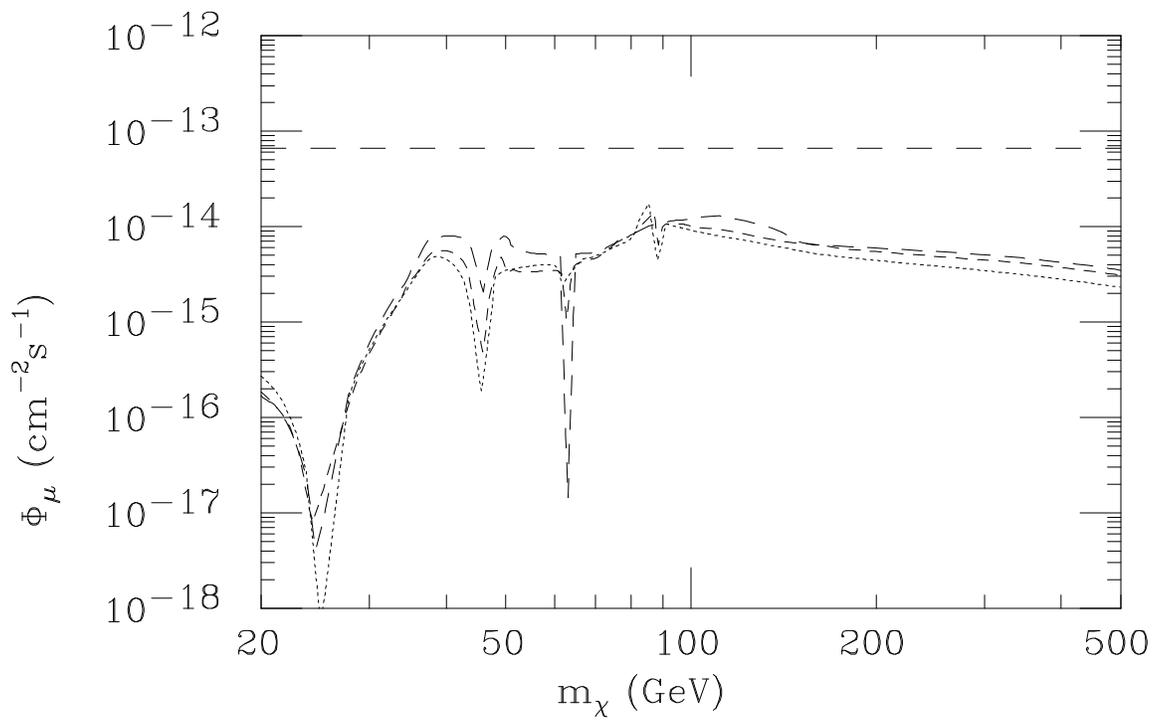
{\bf Figure 13}
Flux $\Phi_\mu$ of the up--going muons as functions of $m_\c$
for $\c$--$\c$ annihilation in the Sun.
The flux is orthogonal to the Sun direction and the
threshold for the muon energy is $E^{\rm th}_{\mu} = 2~\gev$.
The three representative neutralino compositions are
$P = 0.1$ (dotted line), $0.5$ (short--dashed line),
$0.9$ (long--dashed line).
The horizontal line
represents the Kamiokande upper bound
$6.6 \cdot 10^{-14} {\rm cm}^{-2} {\rm s}^{-1}$ (90 \% C.L.) [2].

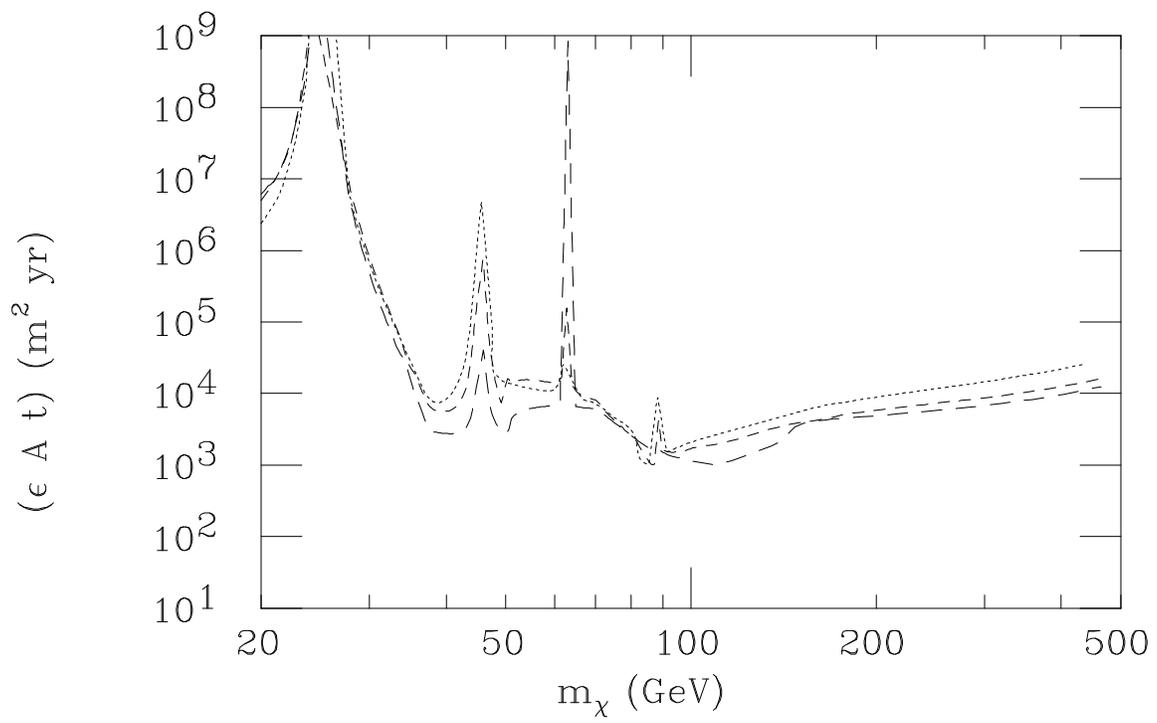
{\bf Figure 14}
Effective exposure $(\epsilon A t)$ necessary to have a $4 \sigma$ effect
with a signal of at least of 4 events,
as a function of neutralino mass $\m$.
The figure refers to
a signal coming from the Sun.
The three representative neutralino compositions are
$P = 0.1$ (dotted line),
$0.5$ (short--dashed line), $0.9$ (long--dashed line).
The half-aperture of the detector is fixed at $5^\circ$ from the Sun
direction.

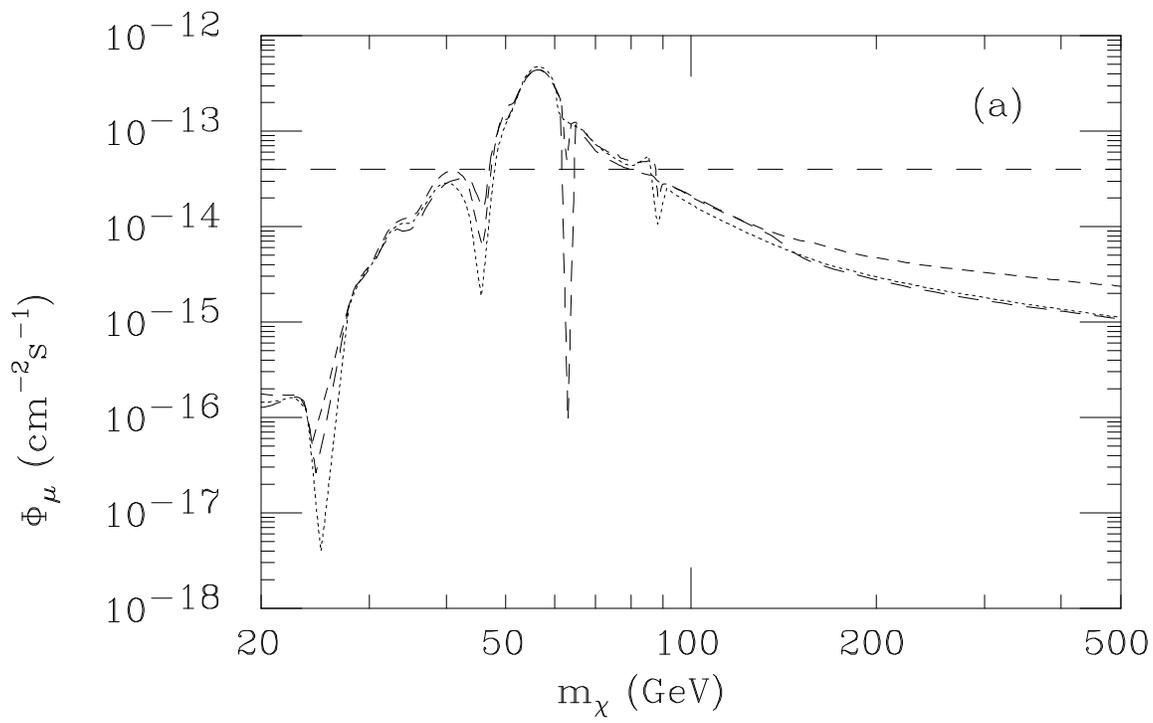
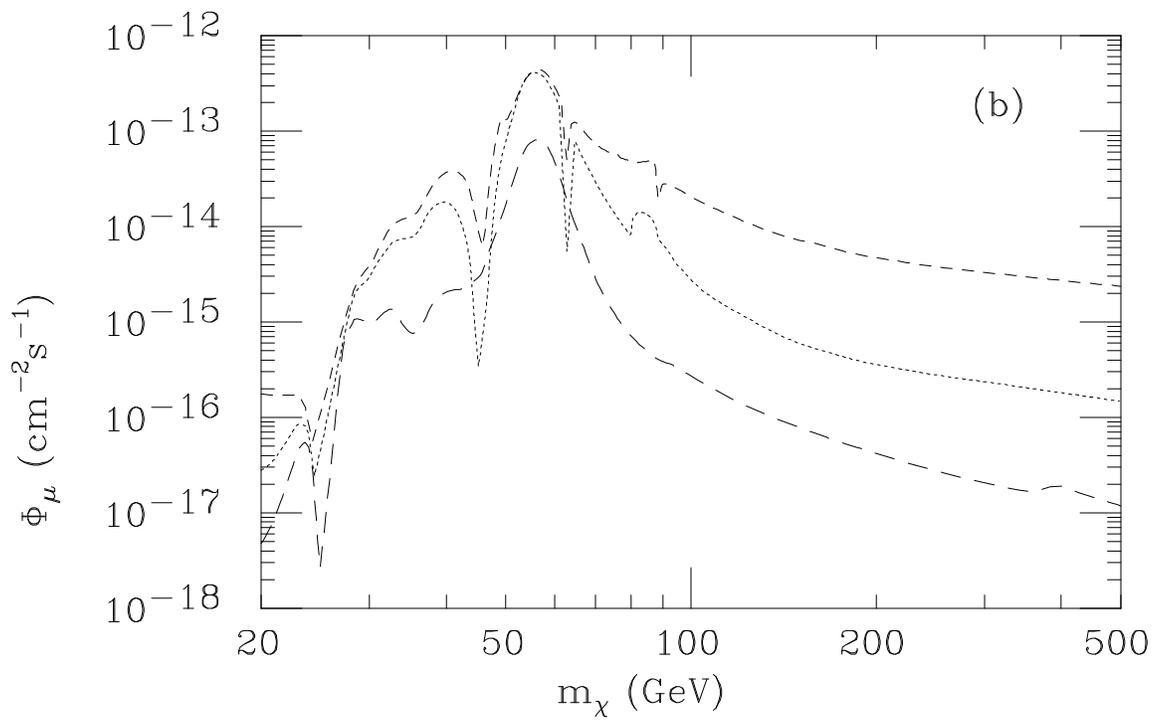
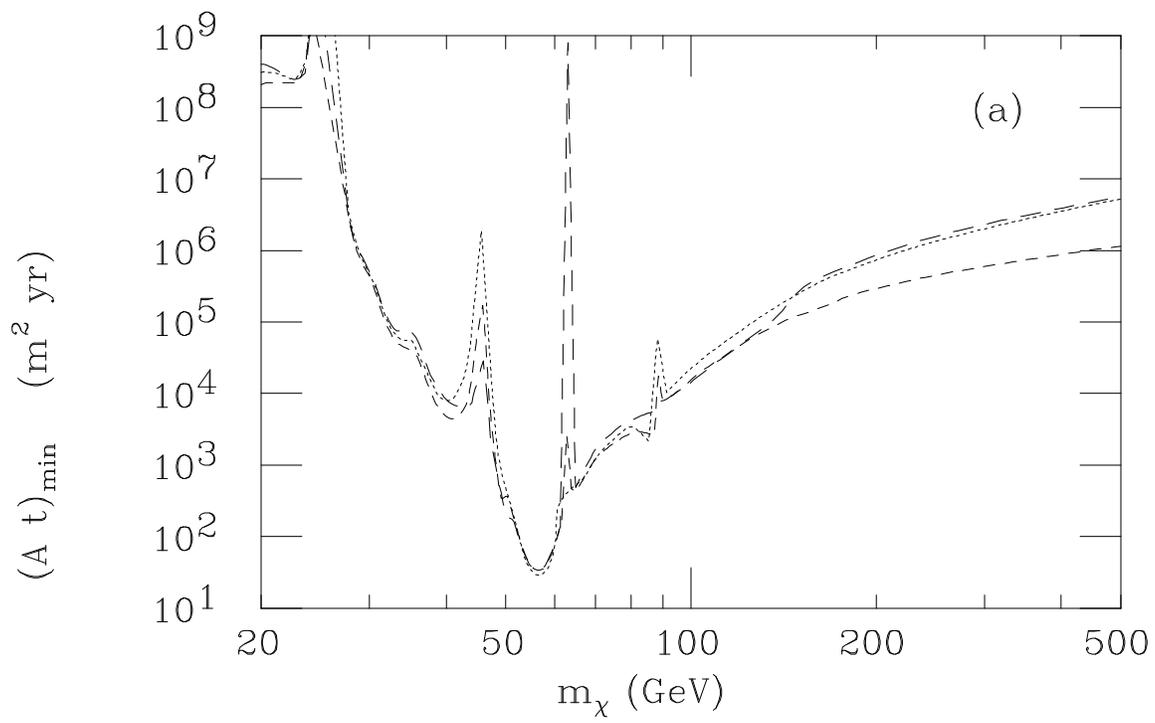
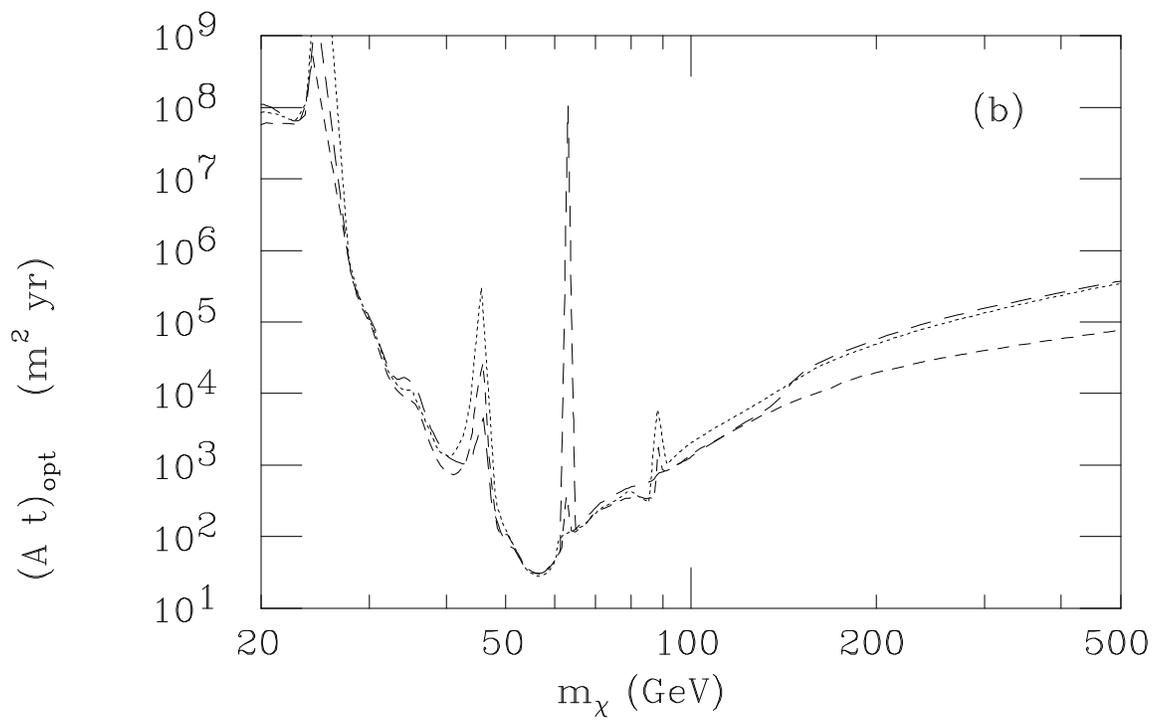
\bye